\documentclass[11pt, a4paper]{article}
\pdfoutput=1
\usepackage{jheppub}
\usepackage{graphicx}
\usepackage{epsfig,amsmath,amsthm}
\usepackage{subfig}
\newcommand{\be}{\begin{equation}}
\newcommand{\ee}{\end{equation}}
\newcommand{\bea}{\begin{eqnarray}}
\newcommand{\eea}{\end{eqnarray}}
\def\bse{\begin{subequations}}
\def\ese{\end{subequations}}

\def\IZ{\relax\ifmmode\hbox{Z\kern-.4em Z}\else{Z\kern-.4em Z}\fi}

\newcommand{\non}{\nonumber \\}
\newcommand{\reef}[1]{(\ref{#1})}
\newcommand{\ie}{{\it i.e.,}\ }
\newcommand{\eg}{{\it e.g.,}\ }
\newcommand{\labell}[1]{\label{#1}}

\def\del{{\partial}} \def\dag{\dagger}
\def\ndel{\displaystyle{\not}\partial}

 \def\hb{{\hat b}}

\def\al{\alpha} \def\bt{\beta}

\newcommand{\Om}{\Omega}
\newcommand{\om}{\omega}
\newcommand{\Lam}{\Lambda}

\def\bi{\begin{itemize}} \def\ei{\end{itemize}}

\def\({\left(} \def\){\right)}
\def\[{\left[} \def\]{\right]}
\title{ \center{(Non) supersymmetric quantum quenches}}

\author{Ling-Yan Hung$^1$, Michael Smolkin$^2$ and Evgeny Sorkin$^3$\\
{\it ${}^1$ Department of Physics, Harvard University,
\\ ~~Cambridge MA 02138, USA} \\
{\it ${}^2$ Perimeter Institute for Theoretical Physics,
\\ ~~31 Caroline Street North, Waterloo,
Ontario N2L 2Y5, Canada} \\
{\it ${}^3$ Department of Physics and Astronomy, University of British Columbia,\\
~~Vancouver, BC V6T 1Z1, Canada} \\

  {\tt\href{mailto:elektron.janethung@gmail.com}{~~elektron.janethung@gmail.com}}\\
 {\tt\href{mailto:msmolkin@perimeterinstitute.ca}{~~msmolkin@perimeterinstitute.ca}}\\
 {\tt \href{mailto:evgeny@phas.ubc.ca}{~~evgeny@phas.ubc.ca}}

}

\abstract{We explore supersymmetric quantum quenches of the mass and
coupling constants in the $\mathcal{N}=1$ supersymmetric vector model
using Hartee-Fock approximation. We find that in the case of 
the free fermionic field,  quench of the mass generates
singularity localized at the instant of sudden change and we point out the essential role of the parity of spacetime. Focusing on a supersymmetric generalization of the $\phi^6$ model and
using stationary phase approximation, we demonstrate that supersymmetry is broken in the asymptotic state that emerges at late times after the quench. Finally, we confirm SUSY breaking in the time-dependent setting by integrating numerically the exact equations of
motion from the instant of the quench into asymptotic regime.  We argue that the breaking of supersymmetry cannot be attributed to the thermal physics since to leading order in the Hartee-Fock approximation the final state is not thermal.}

\begin{document}
\maketitle

\section{Introduction}

%%%%%%%%%%%%%%%%%%%%%%%%%%%%%%%%%%%%%%%%%%%%%%%%%%%%%%%%%%
Research activities towards the understanding of how to characterize and control matter away---especially very far away---from equilibrium 
experience a rapid growth during the past few years. New advances in the experimental front call for the theoretical physics community to rise to a computational challenge, notably in the investigation of 
out-of-equilibrium cold atomic gases, where minute control and precision measurements are now possible within a time-scale where quantum coherence is maintained, see for example \cite{relax}.  

The importance of analytical tools for handling quantum evolution away from equilibrium cannot be overstated.
They are intertwined with the first principles of quantum physics, which underly diverse fundamental phenomena, such as: thermalization of an expanding gas or quantum plasma, the  appearance of new critical behaviour in exotic materials and eventually the dynamics of the early universe. 

Even today, quantum evolution of the many body system remains a hard computational problem that has no general solution or a systematic self-consistent approach. In the presence of interactions only limited number of analytical approaches exist that allows for exact solutions  feasible in certain regimes.  AdS/CFT correspondence \cite{Maldacena:1997re} is one of them. It has provided several tractable examples and insights to the quenched systems, see, for example, \cite{AdSquench} and references therein.

Imposing certain restrictions on the parameters of the system may provide additional
instance when non-equilibrium dynamics admits a self-consistent solution. The adiabatic process is such an example. In that case the external change that perturbs the system out of equilibrium, \eg change in the external magnetic field coupled to a spin chain,  is sufficiently slow compared to the characteristic relaxation time, and full dynamical solution can be constructed perturbatively. 

In the opposite limit, the external environment is assumed to change abruptly at some
point in time, but otherwise remains constant. The process can be approximated by Dirichlet boundary conditions on the field variables at the instant of the sudden change. Studies of this class of dynamics, termed the \emph{quantum quench},  were pioneered in the seminal paper \cite{Cardy1}.  The state of the system right after the quench is highly excited and its subsequent relaxation is of particular interest. So far a lot of results, particularly for integrable models and in low dimensions, such as $1+1$-dimensions, have been obtained \cite{oned}. In many of them, 
equilibrium physics is found to be remarkably well described by the generalized Gibbs ensemble \cite{conjecture}. 

In \cite{Sotiriadis:2010si} Sotiriadis and Cardy paved a way for handling quantum quenches in  higher dimensional interacting field theories. They took advantage of the large-$N$ approximation \cite{Weinberg:1997rv} to investigate quantum evolution of the scalar $O(N)$ vector model whose mass and $\phi^4$ coupling undergo sudden change. To leading order in the large-$N$ expansion this model is effectively free, and the only remnant of interaction is encoded in the time dependent effective mass of the field. In particular, the coupling between modes of different momenta is suppressed and thermalization does not occur. 

The key insight of \cite{Sotiriadis:2010si} rests on two assumptions: (i) The effective mass approaches a stationary value at sufficiently large times and, (ii) Its evolution can be approximated by a jump. These provide  dramatic simplification of the quenched dynamics problem. In \cite{Hung:2012zr} the method was used to study the impact of quantum quenches on the effective potential of a $\phi^6$ field theory model. It was shown that the phase structure is highly sensitive to the details of quantum quench. In particular, new phases, corresponding to extra minima of the effective potential, emerge and the critical value of the coupling constant, beyond which the theory is unstable, is modified. 

Furthermore, symmetry breaking and subsequent symmetry restoration is a particularly interesting realm that can be explored in the context of quantum quenches. Sudden changes obviously break certain symmetries. However, if the system relaxes and thermalizes, it would be desirable to understand if and when symmetry restoration emerges in the process. Time reversal is perhaps the simplest example of broken symmetry that exhibits restoration when the system asymptotically approaches a steady state. 

In this paper we investigate the consequences of quantum quenches on supersymmetric system. Supersymmetry (SUSY), and especially its breaking, play an essential role in phenomenology mainly because of potential relevance to various fundamental aspects of high energy physics that span from the hierachy problem to the cosmological constant and dark matter conundrums. If there is a supersymmetry in Nature, it must be broken and therefore various mechanisms of SUSY breaking attracted attention of theoretical physicists ever since the realm of supersymmetry was discovered. SUSY cannot be broken perturbatively,  since quantum corrections preserve supersymmetry at all orders in the perturbative expansion. However, perturbative analysis breaks down when the system is subject to the quantum quench \cite{Sotiriadis:2010si}, and here we argue that supersymmetric quench of certain field theory model leads to SUSY breaking.

We focus on the simplest supersymmetric extension of the three dimensional $\phi^6$ vector model. In the presence \cite{Bardeen:1984dx,Moshe:2002ra} or absence \cite{Townsend:1975kh} of supersymmetry this model possesses a complex phase diagram. Therefore, it conveniently plays a role of a unique laboratory for studies of various phenomena, \eg big bang cosmology \cite{Craps:2009qc}, Vasiliev's higher spin gravity \cite{Aharony:2011jz} etcetera. To avoid explicit breaking of SUSY we study supersymmetric quantum quenches, \ie parameters of the system are fine-tuned to keep the Hamiltonian supersymmetric before and after the sudden change. Then we follow the approach \cite{Sotiriadis:2010si} and demonstrate that quantum quench leads to the breaking of supersymmetry in this model.  Importantly, the SUSY breaking cannot be attributed to thermal physics as there is no effective temperature that can be assigned to the final state. 

Furthermore, we find quite generally that the structure of the effective potential is modified by the quench. In contrast to the non supersymmetric case \cite{Hung:2012zr}, where competing vacua and phase transitions emerged, in the case of SUSY there is only one stable vacuum in the final state. We support our results by numerical dynamical simulations based on the methods of \cite{Sotiriadis:2010si}.  While there is a qualitative agreement between estimated and numerically calculated asymptotic effective masses, the specific numbers do not match 
as well as in the case of scalar field theory.
However, it is perhaps not too surprising that numerics in the SUSY case do not match analytic computations to the same extent as in the scalar case studied in \cite{Sotiriadis:2010si}. For one thing, the fermionic field obeys first order Dirac equation, whereas scalar field satisfies second order Klein-Gordon equation. Hence, derivative of the fermionic field with respect to time is not continuous across the quench while temporal derivative of the scalar field is smooth, we elaborate on this in concluding remarks section. 

We take advantage of the free field theories to uncover the differences in the response of the scalar and fermionic fields to sharp quantum quenches of the mass parameter. Such theories are Gaussian, and therefore any correlator is given in terms of two point functions that can be expressed as an infinite sum over contributions of instantaneously quenched harmonic oscillators\footnote{In this setup (simple harmonic oscillator) the limit of sudden quenches is a well-posed and exactly solvable problem that does not suffer from divergences suggested by the holographic calculations in \cite{Buchel:2012gw}.}. We find that Hamiltonian density of the fermionic field diverges at the instant of quench, but remains finite otherwise. In odd dimensions this divergence is ultra local, \ie proportional to the delta function and its derivatives are supported at the instant of quench, whereas in even dimensions it is power law $\sim t^{1-d}$, where $d$ is the number of spacial dimensions.
Appearance of singularities at the instant of sharp quench supports observations made in \cite{Buchel:2012gw}, which apparently makes its treatment a very intricate problem in comparison to pure scalar case. Fortunately, in three dimensions this problem can be easily circumvented by adopting dimensional regularization to remove singularities associated with the delta function and its derivatives.

The organization of the paper is as follows: in section \ref{major} we consider quantum quenches of the free Majorana field in three spacetime dimensions and find a general solution, in section \ref{sec:scalar} we investigate supersymmetric quantum quenches of the $\mathcal{N} = 1$ supersymmetric vector model, numerical time-evolution of the effective masses is explored in \ref{sec_dynamics} and concluding remarks are relegated to section \ref{sec:concl}.

\section{Linearly coupled Majorana oscillators}
\label{major}

In this section we study sharp quantum quench of a free Majorana field in 2+1 dimensions. The results obtained here will be used in the next section when we explore a particular supersymmetric model that undergoes (supersymmetric) quantum quench. Based on this simple example we argue that fermionic field responds in a substantially different way to quantum quenches than their scalar counterpart \cite{Sotiriadis:2010si}.  In particular, the expectation value of the fermionic mass term $\langle \bar\psi\psi\rangle$ exhibits singular behaviour at the instant of quench and therefore requires careful consideration. This intricate characteristic of sharp quenches was observed numerically in the context of AdS/CFT correspondence in \cite{Buchel:2012gw}, see also \cite{new} for analytic approach. Here we provide a simple field-theoretic example in favour of this pattern.

Since the properties of Majorana spinors in three space-time dimensions may not be universally known, we briefly recall some of them and explain our notation. Majorana field is given by the condition
 \be
 \psi=C\bar\psi^T~,
 \ee
where $\bar\psi=\psi^{\dag}\gamma^0$, $T$ stands for transpose, and $C$ is the charge conjugation matrix obeying
 \be
 C\gamma^{\mu\,T} C^{-1}=-\gamma^{\mu}~.
 \ee

We use mostly minus signature and construct $\gamma$-matrices out of Pauli matrices, $\gamma^{\mu}=(\sigma_y,-i\sigma_z,i\sigma_x)$. This is the so-called Majorana representation. In this representation all  $\gamma$-matrices are purely imaginary and the Majorana field is real. Indeed, since $\sigma_y\sigma^*\sigma_y=-\sigma$ with $\sigma$ being an arbitrary Pauli matrix, one can check that $C=-\sigma_y$. As a result the Majorana condition reads
 \be
  \psi=C\bar\psi^T=-\sigma_y(\psi^\dag\sigma_y)^T=\psi^*~,
 \ee
Therefore the Fourier expansion of the free Majorana field of mass $\mu_0$ takes the following form
 \be
 \hat\psi(x)=\int {d^2  \vec p \over (2\pi)^2} \sqrt{\mu_0\over \omega_{0p}}\[  \hb_{0 p}\,u_{ 0 p}\, e^{-i\omega_{0p} t+i\vec p\cdot \vec x}+ \hb_{ 0 p}^\dag \,u^*_{ 0 p}\, e^{i\omega_{0p} t-i\vec p\cdot \vec x}\]~,
 \label{MajorFourier}
 \ee
where $\omega_{0p}=\sqrt{\vec p^{\,2}+\mu_0^2}$ and $u_{0p}$ is the on-shell Majorana spinor with positive frequency, \ie it satisfies Dirac equation in Majorana representation
 \be
 (\displaystyle{\not} p-\mu_0)u_{ 0 p}=0\, , \quad p^\mu=(\omega_{0p},p_x,p_y)\, .
 \labell{diraceq}
 \ee
Solving this equation, yields
 \be
 u_{ 0 p}={1\over \sqrt{2\mu_0(p_y+\omega_{0p} )} } 
 \left(\begin{array}{c}p_y+\omega_{0p} \\ p_x+i\mu_0\end{array}\right)~.
 \labell{spinor}
 \ee

Note that there is no summation over the spin indices in the Fourier expansion \reef{MajorFourier}. Indeed, complex conjugation in Majorana representation transforms Dirac spinors with mass $\pm\mu_0$ into each other. In particular, $v_{0p}=u^*_{0p}$ represents positive-frequency spinor with mass $-\mu_0$ or equivalently negative-frequency spinor with mass $+\mu_0$. However, in 3D the spinor space is two dimensional and therefore there are no other independent positive-frequency spinors which are eigenvectors of $\displaystyle{\not} p$ or equivalently solve the Dirac equation.

Normalization of $u_{0p}$ is chosen such that the following relations hold
 \bea
 \bar u_{0p} u_{0p}&=&1=-\bar v_{0p} v_{0p}~, \non
 u^{\dag}_{0p}u_{0p}&=&{\omega_{0p} \over \mu_0}=v^{\dag}_{0p}v_{0p}~,\non
 \bar v_{0p}u_{0p}&=&0= \bar u_{0p}v_{0p}~,\non
 v^{\dag}_{0p}u_{0-p}&=&0~.
 \labell{orthrel}
 \eea

In order to satisfy the standard equal-time anticommutation relations
 \be
 \{\hat\psi_\al(t,\vec x),\hat\psi_{\beta}(t,\vec y)\}=\delta^{(2)}(\vec x - \vec y) \delta_{\al\bt}~,\\
 \labell{commrel}
 \ee
creation and annihilation operators must obey the following anticommutation rules
 \be
 \{\hb_{0p},\hb^{\dag}_{0q}\}=(2\pi)^2 \delta( p -  q)~,
 \labell{commrel2}
 \ee
with all other anticommutators equal to zero. 
 
The quantum quench that we are going to consider here consists of an instantaneous change of the mass from $\mu_0$ to $\mu$ everywhere in space at time $t=0$ . We assume that before the quench the state of the system corresponds to the ground state of the initial hamiltonian $|\Psi_0\rangle$. In addition, the system is kept isolated from the environment immediately before and after the instant $t=0$.
 
Since the theory is free, we deduce that immediately after the quench Majorana field takes the following form
 \be
 \hat\psi(x)=\int {d^2  \vec p \over (2\pi)^2} \sqrt{\mu\over \omega_{p}}\[  \hb_{p}\,u_{ p}\, e^{-i\omega_{p} t+i\vec p\cdot \vec x}+ \hb_{ p}^\dag \,u^*_{ p}\, e^{i\omega_{p} t-i\vec p\cdot \vec x}\]~,
 \ee
where $\omega_{p}=\sqrt{\vec p^{\,2}+\mu^2}$ and $u_p$, $\hb_{p}$, $\hb_{ p}^\dag$ satisfy eqs. \reef{diraceq}-\reef{commrel2} with $\mu_0$ replaced by $\mu$.

Creation and annihilation operators before and after the quench are related by boundary conditions at $t=0$. To uncover these relations, let us examine the Heisenberg equation of motion 
 \be
 \dot{\hat\psi}=i[\hat H, \hat\psi]~.
 \ee
Integrating it in the infinitesimal neighbourhood $\delta$ of the instant $t=0$, and taking the limit $\delta\rightarrow 0$, we deduce that even though the Hamiltonian is only piecewise smooth, the field operator $\hat\psi(x)$ is continuous across the quench. Note, however, that $\dot{\hat\psi}$ exhibits an abrupt jump at $t=0$. Indeed,
 \be
 \Delta\dot{\hat\psi}\equiv\dot{\hat\psi}\big|_{t=0^+}-\dot{\hat\psi}\big|_{t=0^-}=i[\Delta\hat H, \hat\psi|_{t=0}]
 ={i(\mu-\mu_0)\over 2}\int d^2\vec x\,[\hat{\overline\psi}\,\hat\psi, \hat\psi]\big|_{t=0}=i(\mu-\mu_0)\hat{\overline\psi}|_{t=0}~,
 \label{psijump}
 \ee
where in the last equality we used equal-time anticommutation relations \reef{commrel}. 

Form this perspective the behaviour of the Majorana fermion is different from its scalar counterpart since in the scalar case both the field and its time derivative are continuous across the quench. The difference between these two cases emanates from the Heisenberg equations of motion: while in the scalar case it yields the second order Klein-Gordon equation for $\hat\phi$, in the case of Majorana (or Dirac) field $\hat\psi$, it boils down to the standard first order Dirac equation.

Imposing continuity of the field $\hat\psi$ across the quench, leads to the following relations between creation and annihilation operators before and after the quench
\be
\sqrt{\mu\over \omega_{p}}\[  \hb_{p}\,u_{ p}\, + \hb_{ -p}^\dag \,u^*_{ -p}\,\]=\sqrt{\mu_0\over \omega_{0p}}\[  \hb_{0p}\,u_{ 0p}\, + \hb_{ 0-p}^\dag \,u^*_{ 0-p}\,\]~.
\ee
Based on orthogonality relations \reef{orthrel} for $u_{0p}$ and similar relations for $u_{p}$ yields the following Bogoliubov transformation between creation-annihilation operators
\be
\left(\begin{array}{c} \hb_p \\ \hb_{-p}^{\dag}\end{array}\right)=\sqrt{\mu\,\mu_0\over\omega_p\,\omega_{0p}}
\left(\begin{array}{cc} u_p^{\dag} u_{0p} & u_p^{\dag} u_{0-p}^* \\ v_{-p}^\dag v_{0p}^*& v_{-p}^\dag v_{0-p}^{}\end{array}\right)
\left(\begin{array}{c}\hb_{0p} \\ \hb_{0-p}^{\dag}\end{array}\right)~.
\ee

Now we have all ingredients to construct Feynman propagator for the free Majorana field after the quench. In momentum space it takes the following form
\begin{multline}
\langle\Psi_0| \mathcal{T} \{\hat{\psi}_{p \al}(t_1) \, \hat{\overline\psi}_{q \bt}(t_2)\} |\Psi_0\rangle=(2\pi)^2\delta^{(2)}(p+q){\mu^2 \mu_0\over \omega_p^2\omega_{0p}}
\\
\times \Bigg[\theta(t_1-t_2)\Big( e^{-i\omega_p(t_1-t_2)} \bar u_{p\beta}u_{p\alpha}(u_p^\dag u_{0p})(v_p^\dag v_{0p})  
+e^{i\omega_p(t_1-t_2)} \bar v_{-p\beta} v_{-p\alpha}(u_{-p}^\dag v_{0p})(v_{-p}^\dag u_{0p})  
\\
e^{-i\omega_p(t_1+t_2)} \bar v_{-p\beta}u_{p\alpha}(u_p^\dag u_{0p})(u_{-p}^\dag v_{0p})  
+e^{i\omega_p(t_1+t_2)} \bar u_{p\beta} v_{-p\alpha}(v_{-p}^\dag u_{0p})(v_{p}^\dag v_{0p}) \Big)
\\
-\theta(t_2-t_1)\Big( e^{-i\omega_q(t_2-t_1)} \bar v_{q\beta}v_{q\alpha}(u_q^\dag u_{0q})(v_q^\dag v_{0q})  
+e^{i\omega_q(t_2-t_1)} \bar u_{-q\beta} u_{-q\alpha}(u_{-q}^\dag v_{0q})(v_{-q}^\dag u_{0q})  
\\
e^{-i\omega_q(t_1+t_2)} \bar v_{q\beta}u_{-q\alpha}(u_q^\dag u_{0q})(u_{-q}^\dag v_{0q})  
+e^{i\omega_q(t_1+t_2)} \bar u_{-q\beta} v_{q\alpha}(v_{-q}^\dag u_{0q})(v_{q}^\dag v_{0q}) \Big)\Bigg]~,
\label{majorprop}
\end{multline}
where $\mathcal{T}$ denotes time ordering operator for fermions. Notice that the propagator contains terms that break time invariance. In Appendix \ref{major2}, we present an alternative derivation of the above result using prescription suggested in \cite{Sotiriadis:2010si}. 

Various terms in the above expression can be written explicitly using eq. \reef{spinor}. Thus, for instance, we find
 \bea
  \bar u_{p\beta}u_{p\alpha}= -( \bar v_{p\beta}v_{p\alpha})^*&=&{1\over 2\mu}\left(\begin{array}{cc}ip_x+\mu & i(\omega_p-p_y) \\ -i(\omega_p+p_y)& -ip_x+\mu\end{array}\right)_{\bt\al}~,
 \non
 (u_p^\dag u_{0p})(v_p^\dag v_{0p}) &=& {(\omega_p+\omega_{0p})^2-(\mu-\mu_0)^2\over 4\mu\mu_0}~,
 \non
 (u_{-p}^\dag v_{0p})(v_{-p}^\dag u_{0p}) &=& {(\mu-\mu_0)^2-(\omega_p-\omega_{0p})^2\over 4\mu\mu_0} ~,
 \non
 (u_p^\dag u_{0p})(u_{-p}^\dag v_{0p}) &=& (v_{-p}^\dag u_{0p})^*(v_{p}^\dag v_{0p})^*={\mu-\mu_0\over 2\mu_0\mu\sqrt{\om_p^2-p_y^2}}(\mu\,p_y-i\, \om_p \, p_x)~, 
 \eea
However, in all relevant computations we only need to evaluate eq.\reef{majorprop} and its derivatives in the limit $t_1=t_2=t$. Using 
\be
 \bar v_{-p} u_p=-{\mu\, p_y+i\, \om_p \, p_x\over \mu \sqrt{\om_p^2-p_y^2}}\, , \quad \quad
 \bar u_{p} v_{-p}=-{\mu\, p_y-i\, \om_p \, p_x\over \mu \sqrt{\om_p^2-p_y^2}}~,
\ee
yields
 \be
  \langle\Psi_0|  \hat{\overline\psi} \, \psi |\Psi_0\rangle
 =-{\mu\over 2} \int{d^2p\over (2\pi)^2}{\omega_p^2+\omega_{0p}^2-(\mu-\mu_0)^2 \over \omega_p^2\omega_{0p}}
 +(\mu-\mu_0)\int{d^2p\over (2\pi)^2} {p^2\over \omega_p^2\omega_{0p}}\cos(2\om_pt)~.
 \labell{ferloop}
 \ee
The second term in this expression breaks time reversal. Such term is also present in the scalar case, and it was shown in \cite{Sotiriadis:2010si} that its contribution is finite and vanishes in the asymptotic future. However, in our case this term diverges. To isolate the corresponding divergence, let us subtract and add $\cos(2\om_pt)/\om_p$ to the integrand of the second term, then eq.\reef{ferloop} takes the following form
 \begin{multline}
  \langle\Psi_0|  \hat{\overline\psi} \, \psi |\Psi_0\rangle
 =-{\mu\over 2} \int{d^2p\over (2\pi)^2}{\omega_p^2+\omega_{0p}^2-(\mu-\mu_0)^2 \over \omega_p^2\omega_{0p}}
 +(\mu-\mu_0)\int{d^2p\over (2\pi)^2} {p^2-\omega_p\omega_{0p}\over \omega_p^2\omega_{0p}}\cos(2\om_pt)\\-(\mu-\mu_0){\sin(2|\mu| t)\over 4\pi t}+{\mu-\mu_0\over 4}\,\delta(t)~,
  \labell{Floop}
 \end{multline}
where we used the following identity
\be
 \int{d^2p\over (2\pi)^2} {\cos(2\om_pt)\over \om_p}=\int_{|\mu|}^\infty {d\om_p\over 2\pi} \cos(2\om_pt)= -{\sin(2|\mu| t)\over 4\pi t}
 +{1\over 4}\delta(t)~.
 \label{iden}
\ee
 
Delta function is not the only singularity of eq.\reef{Floop}. Its first term exhibits linear divergence. However, in dimensional regularization these singularities disappear as $d\to2$. For instance, eq.\reef{iden} becomes
\begin{multline}
 \int  \frac{d^d p}{(2 \pi)^d} {\cos(2\om_pt) \over \om_p} ={1\over 2^{d-1}\pi^{d/2}\Gamma(d/2)}\int_{|\mu|}^{\infty} (\om_p^2-\mu^2)^{d-2\over 2} \cos(2\om_pt) d\om_p 
 \\
 = -{1\over 2^{d}\pi^{d-1\over2}}\Big({t\over |\mu|}\Big)^{1-d\over 2} Y_{1-d\over 2}(2|\mu| t) \label{idnt}~.
 \end{multline}
Therefore the integral is finite and decays as $t^{-d/2}$ in the asymptotic future defined by $t\gg\mu^{-1}$. Remarkably, there is a significant difference between odd and even $d$ for $t\ll |\mu|^{-1}$ in the above expression
\be
\label{odd_even}
  \int  \frac{d^d p}{(2 \pi)^d} {\cos(2\om_pt) \over \om_p} \underset{t\mu\to 0}{\longrightarrow} 
  \left\{\begin{array}{c} {(-1)^{d-1\over 2}\Gamma\big({d-1\over 2}\big)\over 2^d\pi^{d+1\over 2} }~ t^{1-d} \quad \text{odd $d$\,,} \\
   \\ {\Gamma\big({1-d\over 2}\big)\over 2^d\pi^{d+1\over 2} }~ |\mu|^{d-1} \quad\quad\quad~~  \text{even $d$\,.} \end{array}\right.
\ee
In particular, for odd $d$, \ie even dimensional space-time, eq.\reef{idnt} diverges as $t|\mu|\to 0$. Of course, strictly speaking the above computation cannot be taken at face value in general $d$ since the space of Dirac spinors is not the same in various dimensions. However, it suggests that in even dimensional space-time the limit of infinitely sharp quench may require a refined analysis in the presence of fermions \cite{Buchel:2012gw}, we leave investigation of this for future work.

In our case $d$ is even, and therefore either implicitly assuming appropriate regularization scheme or by noting that $\delta(t)=0$ in the region of our main interest ($t|\mu|\to\infty$), we drop the last term in eq.\reef{Floop}. The second term in eq.\reef{Floop} is now finite. It breaks time reversal and decreases in time. More specifically using the stationary phase method, one can show that for large times it decays as $1/t$. Hence, for $t\gg\mu^{-1}$ only first term contributes
 \be
  \langle\Psi_0|  \hat{\overline\psi} \, \psi |\Psi_0\rangle\big|_{t\gg\mu^{-1}}
 =-{\mu\over 2} \int{d^2p\over (2\pi)^2}{\omega_p^2+\omega_{0p}^2-(\mu-\mu_0)^2 \over \omega_p^2\omega_{0p}}~.
 \labell{infmajor2p}
 \ee

Now  using the definition of $\mathcal{T}$, commutation relations \reef{commrel} and the fact that $\hat\psi$ satisfies the Dirac equation, yields the standard Green's equation that Feynman propagator must satisfy 
 \be
(i\displaystyle{\not}\del_{x_1}-\mu) \langle\Psi_0| \mathcal{T} \{\hat{\psi}_{\al}(x_1,t_1) \, \hat{\overline\psi}_{ \bt}(x_2,t_2)\} |\Psi_0\rangle=i\delta(t_1-t_2)\delta^{(2)}(\vec x - \vec y) \delta_{\al\bt}~,
 \ee
We verified that our final expression in eq.\reef{majorprop} indeed satisfies this identity. In particular, it follows that
 \be
 \langle\Psi_0|  \hat{\overline\psi} \,i \displaystyle{\not}\del \,\psi |\Psi_0\rangle = \mu \langle\Psi_0|  \hat{\overline\psi} \, \psi |\Psi_0\rangle~.
\label{major2p}
 \ee

\section{Quantum quenches of $\mathcal{N}=1$ SUSY in 3D}
\label{sec:scalar}

We turn now to explore the impact of quantum quenches on supersymmetry. This problem is particularly interesting in light of our findings in the previous section where we argued that response of the fermionic field to sudden changes in the parameters of the theory is substantially different from the scalar case. Unfortunately, quantum quenches in the presence of interactions is an intricate field theory problem without general self-consistent approach. Therefore we resort to studies of the simplest supersymmetric extension of the $O(N)$ $\phi^6$ model \cite{Bardeen:1984dx} using techniques developed by S.Sotiriadis and J.Cardy in\footnote{Sharp quenches of $\phi^6$ model without supersymmetry were studied in \cite{Hung:2012zr}, see also \cite{Das:2012mt} for quenches of $O(N)$ nonlinear sigma model.} \cite{Sotiriadis:2010si}.  Their method is based on certain assumptions that make the problem of quantum quenches tractable in the large-$N$ limit. Of course, the final results are only reliable provided that these assumptions indeed hold, therefore we test our conclusions numerically in the next section. We find that asymptotic state is stationary, but not thermal and demonstrate that supersymmetry in this state is broken.  

The model that we study is given by $\mathcal{N}=1$  SUSY consisting of an $N$-component real scalar field $\phi$ and an $N$-component, two-component Majorana spinor $\psi$. The corresponding action takes the following form\footnote{The superspace representation of the model can be found in, \eg \cite{Gates:1983nr,Moshe:2002ra}.}
 \bea
 S(\phi,\psi)&=&{1\over 2} \int d^3x \big[\del_\mu\phi\del^\mu\phi-\mu_0^2\phi^2+\bar\psi(i\ndel-\mu_0)\psi
 \non
 &-&2\,{g_0\mu_0 \over N}(\phi^2)^2-{g_0^2\over  N^2}(\phi^2)^3-{g_0\over N}\phi^2(\bar\psi\cdot\psi)
 -2\,{g_0\over N}(\phi\cdot\bar\psi)(\phi\cdot\psi)\big]~.
 \labell{action}
 \eea
Our spinor notation is explained in the previous section. 

At $t=0$ we instantaneously change the mass from $\mu_0$ to $\mu$ and the coupling constant from $g_0$ to $g$. As before, for simplicity we assume that initially $g_0=0$, \ie there is no interaction before the quench and the system is prepared in the ground state of the corresponding free hamiltonian  $|\Psi_0\rangle$.  

Since parameters of the theory are changed abruptly rather than adiabatically, one needs to resort to a well-known Keldysh-Schwinger, or in-in, formalism for non-equilibrium quantum systems. In this approach one needs to impose the boundary conditions only at $t=t_i$, and in our case they are such that the initial state at $t_i=0$ is identical to $|\Psi_0\rangle$. 

In particular, the expectation value of an arbitrary operator $\mathcal{\hat O}(t)$ is given by
 \be
 \langle \Psi_0 | \mathcal{ \hat O}(t)   |\Psi_0\rangle
 =\int_{CTP} D\eta \, \mathcal{ \hat O}(t) \, e^{i S(\eta)}~,
 \ee
where for brevity $\eta$ collectively denotes scalar and Majorana fields and the following notation is used to designate the closed-time-path (CTP) integral measure  
 \be
 \int_{CTP} D\eta=\int D\eta_i \, \Psi_0(\eta_i)\int D \tilde\eta_i  \, \Psi_0^*(\tilde\eta_i)
 \int_{\eta_i}^{\tilde\eta_i}D\eta ~, 
 \ee
where $\eta_i$ and $\tilde\eta_i$ denote the values of the fields at the end points of the time contour, whereas $\Psi_0(\eta_i)=\langle \eta_i|\Psi_0\rangle$ and similarly for the complex conjugate $\Psi_0^*(\tilde\eta_i)$.

In what follows we are going to use the large $N$ (or equivalently Hartee-Fock) approximation in order to explore the evolution of the effective mass associated with the real scalar and Majorana fields. We start from noting that the action is quadratic in $\psi$ and hence the Majorana field can be easily integrated out, and the action takes the following form
 \begin{multline}
 S(\phi)={1\over 2} \int d^3x \big[\del_\mu\phi\del^\mu\phi-\mu_0^2\phi^2-2\,{g_0\mu_0 \over N}(\phi^2)^2-{g_0^2\over  N^2}(\phi^2)^3\big]~.
 \\
 -i{N-1\over 2}\text{Tr}\log\big(i\ndel-\mu-{g\over N}\phi^2\big)-{i\over 2}\text{Tr}\log\big(i\ndel-\mu-3 {g\over N}\phi^2\big) ~.
  \end{multline}
Using now the following identity\footnote{We keep CTP label in the path integral over $\rho$ and $\lambda$ to emphasize that the delta-function is inserted at each point of the Keldysh-Schwinger contour. Obviously there are no boundary conditions associated with $\rho$ and $\lambda$. Note also that the equalities hold up to irrelevant multiplicative constant.}
\be
\mathbb{I} = \int_{CTP} D\rho\,\delta(\phi^2-N\rho)=\int_{CTP} D\rho D\lambda ~e^{-{i\over 2}\int d^3x \,\lambda(\phi^2-N\rho)}~,
\label{identity}
\ee
we can rewrite the path integral over the scalar field $\phi$ as follows
 \be
 \langle \Psi_0 | \mathcal{ \hat O}(t)   |\Psi_0\rangle
 = \int_{CTP} D\phi \int D\rho D\lambda\, \mathcal{ \hat O}(t) \, e^{i S(\phi,\rho,\lambda)}~,
 \ee
where 
\begin{multline}
 S(\phi,\rho,\lambda)=N\int d^3x\[{\lambda\,\rho\over 2}-{g^2 \over 2}\rho^3-g\, \mu\, \rho^2\]-{1\over 2}\int d^3x \phi(\square+\mu^2+\lambda)\phi
 \\
 -i{N-1\over 2}\text{Tr}\log\big(i\ndel-\mu-g\rho\big)-{i\over 2}\text{Tr}\log\big(i\ndel-\mu-3 g\rho\big) ~.
  \end{multline}
Performing the Gaussian integral over $\phi$ yields
 \be
 \langle \Psi_0 | \mathcal{ \hat O}(t)   |\Psi_0\rangle
 =\int_{CTP} D\rho D\lambda \, \mathcal{ \hat O}(t) \, e^{i N S_{eff}(\rho,\lambda)}~,
 \label{pathint}
 \ee
with
 \begin{multline}
 S_{eff}(\rho,\lambda)=\int d^3x\[{\lambda\,\rho\over 2}-{g^2 \over 2}\rho^3-g\, \mu\, \rho^2\]+{i\over 2}\text{Tr}\log(\square+\mu^2+\lambda)
 \\
 -{i\over 2}{N-1\over N}\text{Tr}\log(i\ndel-\mu-g\rho)-{i\over 2N}\text{Tr}\log(i\ndel-\mu-3 g\rho) ~.
 \label{susyact}
 \end{multline}
Since we are interested in the large $N$ limit, we drop the last term in the second line and replace $N-1\sim N$. The boundary conditions are now encoded in the functional traces that explicitly depend on the integration parameters $\lambda$ and $\rho$.  The latter, of course, makes their evaluation extremely difficult.

In what follows we are interested to study the evolution of the effective mass of the fields for $t>0$. When $N\to\infty$ (with $g$ and $\mu$ fixed), the effective mass can be evaluated using the stationary phase approximation. Indeed, in this limit the right hand side of \reef{pathint} is dominated by the field configuration that minimizes \reef{susyact}, \ie solves the corresponding classical equations of motion derived from $S_{eff}(\rho,\lambda)$
 \bea
 m_\phi^2&\equiv&\mu^2+\bar\lambda=\mu^2+4 g \mu \bar\rho+3 g^2\bar\rho^2-g  \int {d^2 p \over (2\pi)^2} \, \text{tr}\,\tilde G_\psi(t,t;p)
 ~,\non
 \bar\rho&=&\int {d^2 p \over (2\pi)^2} \, \tilde G_\phi(t,t;p)~,
 \label{susygap}
 \eea
 where ``$\text{tr}$" denotes the trace over spinor indices,
 $m_\phi^2$ as defined above corresponds to the effective mass of the
 scalar field, while $\tilde G_\psi(t_1,t_2;p)$ and $\tilde
 G_\phi(t_1,t_2;p)$ represent  the leading-$N$ momentum space  two point
 correlation functions of the Majorana and scalar fields respectively.
 Note that $\tilde G_\psi(t_1,t_2;p)$ depends on
 \be
 m_\psi\equiv\mu+g\bar\rho~,
 \ee 
which plays a role of the effective mass of the Majorana field.
 
To explore supersymmetry breaking we consider the following order parameter
 \be
 m_\phi^2-m_\psi^2=2 g m_\psi\bar\rho- g \int {d^2 p \over (2\pi)^2} \text{tr}\, \tilde G_\psi(t,t;p)~.
 \label{orderparam}
 \ee
The right hand side of the above equation must vanish in order to
preserve supersymmetry. Unfortunately, it is difficult to analyze it in full generality since exact solution to the gap equations \reef{susygap} is out of reach. Hence, we resort to approximation
proposed in \cite{Sotiriadis:2010si}, \ie we assume that as $t\to\infty$\footnote{Here and in what follows $t\to\infty$ means that time is much larger than any other length scale in the problem.}, $m_\phi$ and $m_\psi$ approach stationary values $m_\phi^*$ and $m_\psi^*$ respectively, and at late times large compared to the duration of the transients, their evolution can be approximated by a jump. Then the  two point correlation function $\tilde G_\phi(t_1,t_2;p)$ (or $\tilde G_\psi(t_1,t_2;p)$) is approximately the same as the propagator in the massive scalar (or fermion) free field theory in which the mass parameter is instantaneously changed from $\mu_0$ to $m_{\phi}^*$ (or $m_\psi^*$),\ie
 \bea
 \label{free_corr}
 \tilde G_\phi(t_1,t_2;p)&\simeq& G_\phi(t_1,t_2;p;\mu_0, m^*_\phi)~, 
 \non
 \tilde G_\psi(t_1,t_2;p)&\simeq& G_\psi(t_1,t_2;p;\mu_0, m_\psi^*)~,
 \labell{approx}
 \eea
where as shown in \cite{Sotiriadis:2010si}
 \be
 G_\phi(t_1,t_2;p;\mu_0, m^*_\phi)={(\tilde\om_p^*-\tilde\om_{0p})^2\over 4\tilde\om_p^{*2}\tilde\om_{0p}}\cos\tilde\om_p^*(t_1-t_2)
 +{\tilde\om_p^{*2}-\tilde\om_{0p}^2\over 4\tilde\om_p^{*2}\tilde\om_{0p}}\cos\tilde\om_p^*(t_1+t_2)
 +{1\over 2\tilde\om_p^*} e^{-i\tilde\om_p^*|t_1-t_2|}~.
 \label{phicorr}
 \ee 
with  $\tilde\omega_{p}^*=\sqrt{\vec p^{\,2}+m_\phi^{*2}}$ and  $\tilde\omega_{0p}=\sqrt{\vec p^{\,2}+\mu_0^2}$, whereas for Majorana field $G_\psi(t_1,t_2;p;\mu_0, m_\psi^*)$ is given by eq.\reef{majorprop} with $\mu$ there substituted by $m_\psi^*$. These relations are expected to be asymptotically exact. Their validity is scrutinized in the next section, where we present numerical studies of the full time evolution of the  effective masses without implementing assumptions about asymptotic stationarity and fast relaxation.

Note that the second term on the right hand side of eq.\reef{phicorr} as well as terms in the third and last lines of eq.\reef{majorprop} break time invariance, however as argued in the previous section their contribution vanishes in the limit $t_1=t_2=t\gg m_\phi^{*-1},m_\psi^{*-1}$, and hence we drop them in what follows. As a result, asymptotically eq.\reef{orderparam} becomes
 \be
 m_\phi^{*\,2}-m_\psi^{*\,2}=2 g m_\psi^*\int {d^2 p \over (2\pi)^d} G_\phi(t,t;p;\mu_0,m_\phi^*)- g \int {d^2 p \over (2\pi)^2} \text{tr} \, G_\psi(t,t;p;\mu_0,m_\psi^*)~.
 \ee
Now it follows form eq.\reef{infmajor2p} that for $t\gg m_\psi^{*-1}$, the following relation holds
\begin{multline}
\text{tr} \, G_\psi(t,t;p;\mu_0,m_\psi^*)= m_\psi^* \, {\omega_p^{*2}+\omega_{0p}^2-(m_\psi^*-\mu_0)^2\over  2\omega_p^{*2}\omega_{0p}}
\\
=2m_\psi^*G_\phi(t,t;p;\mu_0,m_\psi^*)-m_\psi^* {(m_\psi^*-\mu_0)^2\over  2\omega_p^{*2}\omega_{0p}}\,.
\end{multline}
Hence,
 \bea
 m_\phi^{*\,2}-m_\psi^{*\,2}&=&2 g m_\psi^*\int {d^2 p \over (2\pi)^2} \bigg( G_\phi(t,t;p;\mu_0,m_\phi^*)- G_\phi(t,t;p;\mu_0,m_\psi^*) \bigg)
 \non
 &+&{g\, m_\psi^*\over 4\pi} {(m_\psi^*-\mu_0)^{2}\over (m_\psi^{*2}-\mu_0^2)^{1/2}}\arccos\bigg( {\mu_0\over |m_\psi^*|} \bigg)~.
 \labell{order}
 \eea
The first thing to note about this expression is that each loop integral linearly diverges, however the divergent contributions cancel between scalar and fermionic loops exhibiting a supersymmetric nature of the underlying model.  

Let us now ask if any supersymmetric solution to the gap equations \reef{susygap} exists within our approximation. From eq.\reef{order} such solution (if exists) must satisfy $m_\phi^{*}=m_\psi^{*}=\mu_0$. It corresponds to a supersymmetric state that emerges as $t\to\infty$ and is characterized by the same mass parameter $\mu_0$ as before the quench. Plugging this solution back into the definition of the Majorana mass $m_\psi=\mu+g\bar\rho$ results in the following constraint
 \be
 \mu_0=\mu+g\int {d^d p \over (2\pi)^d} {1\over 2\sqrt{p^2+\mu_0^2}}\underset{d\rightarrow 3}{=}\mu-{g\over 4\pi}|\mu_0| ~.
 \ee
This constraint represents a family of supersymmetric solutions, but it cannot be satisfied for all values of $\mu, \mu_0$ and $g$ and therefore generically SUSY is broken in the final state\footnote{Strictly speaking this constraint should hold only approximately since it was derived within certain approximation scheme.}. For instance, one such fine-tuned supersymmetric solution is given by $m_\phi^{*}=m_\psi^{*}=\mu_0>0, \,g=-4\pi$ and $\mu=0$. However, as we demonstrate below it represents an inflection point rather than (local) minimum of the effective potential, and hence does not correspond to a (meta)stable phase of the theory. Numerical calculations of the next section support this conclusion.

\begin{figure}[t!]
\begin{center}
\includegraphics[scale=0.95]{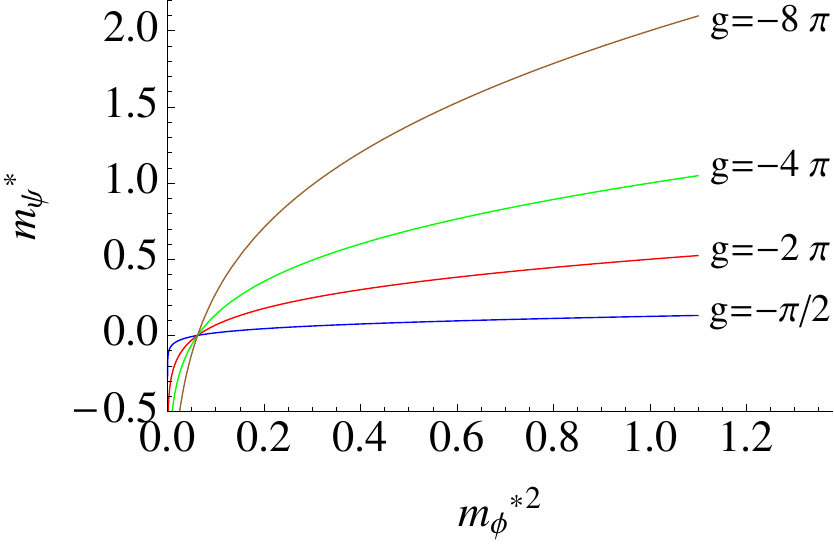}
\caption{The stationary fermion mass, $m_\psi^*$,  as a function of
  the scalar mass
  $m_\phi^*$ for $\mu_0=1$, $\mu=0$ and
  a set of values of the coupling constant $g$. }
\label{fig:mf_ms}
\end{center}
\end{figure}

Implementing approximation \reef{approx} at the level of the gap equations \reef{susygap}, we find that stationary values $m_\phi^*$ and $m_\psi^*$ satisfy the following equations
 \bea
 m_\psi^{*}&=&\mu+g\bar\rho~,
 \label{mpsi}
 \\
  \bar\rho&=&\lim_{t\to\infty}\int {d^2 p \over (2\pi)^2} \, \tilde G_\phi(t,t;p)=-{1\over 4\pi}\bigg(\mu_0+{1\over 2}\sqrt{m_\phi^{*2}-\mu_0^2}\,\arccos(\mu_0/ |m_\phi^*|)\bigg)~,
  \labell{susygap2}
 \\
 m_\phi^2&=&\mu^2+4 g \mu \bar\rho+3 g^2\bar\rho^2
 +{g \, m_\psi^*\over 2\pi}\bigg(\mu_0+{1\over 2}\sqrt{m_\psi^{*2}-\mu_0^2}\,\arccos (\mu_0/ |m_\psi^*|)\bigg)
 \non
 &&\quad\quad\quad\quad\quad\quad\quad\quad\quad\quad\quad\quad~
  +{g\, m_\psi^*\over 4\pi} {(m_\psi^*-\mu_0)^{2}\over (m_\psi^{*2}-\mu_0^2)^{1/2}}\arccos\bigg( {\mu_0\over |m_\psi^*|} \bigg)
 ~,
   \labell{susygap3}
  \eea
where dimensional regularization has been used to regulate the divergent loop integral in the definition of $\bar\rho$. Solutions to these equations describe the stationary points of the effective potential and correspond to various  phases of the theory. Typical plots of $m_\psi^*$ as a function of $m_\phi^{*2}$ for $\mu_0=1$, $\mu=0$ and various values of the coupling constant $g$ are shown in figure \ref{fig:mf_ms}. 

In the next subsection we construct the effective potential and explore its behaviour as a function of parameters of the theory. 

\subsection*{Effective potential}

To analyze the phase structure of the model let us consider the effective potential of the theory as $t\rightarrow\infty$. To leading order in the $1/N$ expansion it is given by eq.\reef{susyact} evaluated for constant $\bar\rho$ and $\bar\lambda$ (or $m_\phi^{*2}$). Up to irrelevant constant it takes the following form
 \bea
 V_{eff}(\bar\rho,m_\phi^{*2})&=&{\mu^2\over 2}\bar\rho+g\mu\bar\rho^2+{g^2 \over 2}\bar\rho^3-{m_\phi^{*2}\,\bar\rho\over 2}
 +{1\over 2} \int_0^{m_\phi^{*2}} dm^2 \int {d^2 p \over (2\pi)^2} \, G_\phi(t,t;p;\mu_0,m)
 \non
 &&-{1\over 2}\int_0^{\mu+g\bar\rho} dm \int {d^2 p \over (2\pi)^2} \text{tr} \, G_\psi(t,t;p;\mu_0,m)~.
 \labell{effpot}
 \eea
Varying this effective potential with respect to $\bar\rho$ and $m_\phi^{*2}$ reproduces the saddle point equations \reef{susygap}. Now we can use eq.\reef{susygap2} to eliminate the auxiliary field $\bar\rho$ and express  $V_{eff}$ in terms of variational parameter $m_\phi^{*}$ only. The value of the physical mass $m_\phi^{*}$ corresponds to the minimum of the resulting $V_{eff}$. 

The role of the effective potential within Hartee-Fock approximation resembles the free energy in thermodynamics that must be minimized at equilibrium with respect to any unconstrained internal variable for a closed system at constant temperature and volume. In particular, the system is stable if and only if the free energy as a function of unconstrained variables is bounded from below. In our case for $m_\phi^{*}\gg\mu_0,\mu$, we have
\be
 V_{eff}\simeq{1024-12 g^2+(g^2)^{3/2}\over 98304} m_\phi^{*3} >0~,
\ee
and therefore as expected in supersymmetric theory the potential is always bounded from below and the system is stable.

Let us analyze the effective potential as a function of $g$ when the only dimensional parameter $\mu$ is set to zero. In this case the theory right after the quench is conformal since to leading order in $1/ N$ all anomalous dimensions vanish. In appendix \ref{emtrace} we explicitly verify that expectation value of the trace of supersymmetric energy momentum tensor indeed vanishes in the quenched state.

\begin{figure}[t]
\begin{center}
\includegraphics[scale=1]{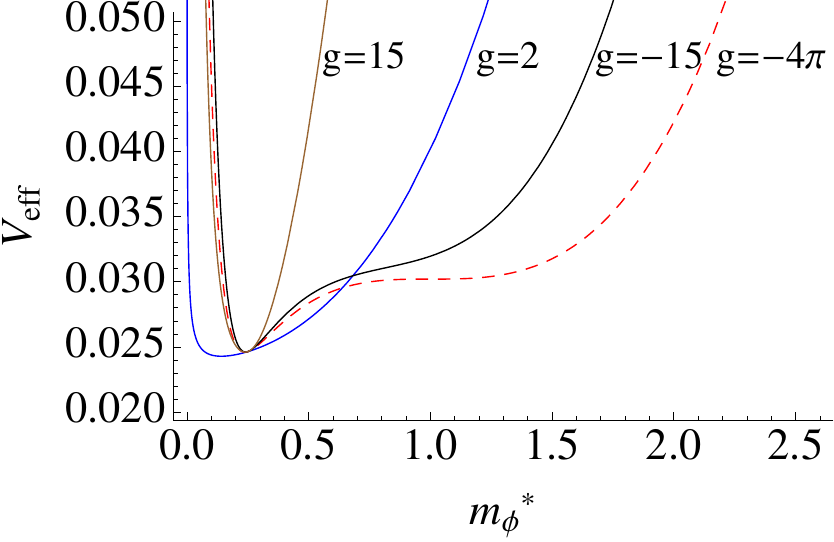}
\caption{Effective potential \reef{effpot} as a function of $m_\phi^*$
  for several values of the coupling $g$, $\mu_0=1$ and $\mu=0$. }
\label{fig:phases}
\end{center}
\end{figure}

The characteristic shape of the effective potential \reef{effpot} for several choices of the coupling constant $g$ is shown in figure \ref{fig:phases}. The only supersymmetric solution of  the gap eqs. \reef{mpsi}-\reef{susygap3} for this choice of $\mu$ is given by  $m_\phi^{*}=m_\psi^{*}=\mu_0$ and $g=-4\pi$. However, as shown on the plot, this solution corresponds to an inflection point rather than to a minimum of $V_{eff}$, and therefore it does not represent a stable phase of the theory. Hence, we conclude that in this case supersymmetry is always broken in the final state. 

%Remarkably, in the unquenched case \cite{Bardeen:1984dx,Moshe:2002ra}, $\mu=0,g=-4\pi$ corresponds to a point on the phase diagram where scale invariance is spontaneously broken. 

Table \ref{tab} presents the masses $m_\phi^*$ and $m_\psi^*$ for a select set of values of the coupling constant $g$. 
\begin{table}[ht]
\centering 
\begin{tabular}{c c c } 
\hline\hline 
 & $m_\phi^*$ & $m_\psi^*$ \\ [0.5ex] 
\hline 
$g=5$& $0.199$ & $0.05$\\ [0.5ex] 
\hline 
$g=-10$& $0.234$& $-0.03$\\ [0.5ex]
\hline
$g=-4\pi$& $0.239$ & $-0.025$ \\[0.5ex]
\hline
\end{tabular}
\caption{ Masses of the particles for various values of the coupling constant $g$, $\mu_0=1$ and $\mu=0$.}
\label{tab}  
\end{table}
Remarkably, susy breaking in this case cannot be attributed to thermal physics since as argued in \cite{{Sotiriadis:2010si},Hung:2012zr} the theory is integrable to leading order in $1/N$, and therefore the final state cannot be described by an emergent effective temperature.

%%%%%%%%%%%%%%%%%%%%%%%%%%%
\section{Dynamical evolution of the effective masses.}
\label{sec_dynamics}
%%%%%%%%%%%%%%%%%%%%%%%%%%%

We have analysed the phase structure of the model,
assuming  that the
effective masses relax to their asymptotic values at late times, after
an instantaneous initial transient. The full time evolution will require
solving the gap equation (\ref{susygap}) to the future of the quench
$t>0^+$. While difficult to achieve in general, the problem becomes
tractable under the assumption of short transient and subsequent evolution
described by the free propagators (\ref{free_corr}), with
time-dependent effective masses \cite{Sotiriadis:2010si}. In this
section we test this hypothesis and solve the dynamical evolution problem by integrating the effective masses in
time. 

To leading order in the large-$N$ expansion the model is effectively free. In particular, it  can be decomposed into
a set of independent harmonic modes in momentum space that are coupled via the time-dependent effective masses
of scalar and Majorana fields, see eq.\reef{susygap} and discussion thereafter
\be
\label{meff_t}
m_{\phi}^2(t) = \mu^2+4g\mu\bar\rho+3 g^2 \bar\rho^2+ g \, \bar\rho_\psi, ~~~~~~~
m_\psi(t)=\mu + g \, \bar\rho
\ee
where
\bea
\label{rhos}
\bar\rho&=&{1\over N}\langle \phi^2\rangle=\int {d^2 p \over (2\pi)^2} \, \tilde G_\phi(t,t;p),
\nonumber\\
\bar\rho_\psi&=&{1\over N}\langle \bar\psi\psi\rangle=-\int {d^2 p \over (2\pi)^2} \text{tr}\, \tilde G_\psi(t,t;p).
\eea
We require that $\phi$ and $\psi$ are continuous across the instant of quench,  it follows that at $t=0^+$ the effective masses are given by
\bea
\label{id}
m_{\phi}^2(0^+) &=&\mu^2+4 g \mu \({-\mu_0\over 4\pi}\)+3 g^2 \({-\mu_0\over 4\pi}\)^2-2 g \mu_0
\({-\mu_0\over 4\pi}\) \, ,
\non
m_\psi(0^+)&=&\mu+g \({-\mu_0\over 4\pi}\),
\eea
where loop integrals were evaluated in the free field theory, and dimensional regularization was used to handle divergences.

The time-evolution of the effective masses can be formulated as initial value problem. 
Indeed, as argued in \cite{Sotiriadis:2010si} the equations of
motion governing the modes with spatial momentum $p$, can be solved by introducing the following ansatz
\be
\label{y_Omega}
\hat\Psi_p(t) \sim \frac{1}{\sqrt{2\, \Omega_{\Psi }(t)}} \, \exp\(-i\, \int^t_0
\Omega_{\Psi} (t') dt' \),
\ee 
where $\hat\Psi=(\hat\phi,\hat\psi)$ denotes collectively the scalar and Majorana field operators. Substituting this ansatz into the Heisenberg equation of motion obeyed by each momentum mode separately,
 \be
 \ddot{\hat\Psi}_p(t)+\omega_\Psi^2(t)\hat\Psi_p(t)=0,
 \ee
where $\omega_\Psi^2(t)=m_{\Psi}(t)^2+p^2$, yields the following nonlinear equation for $\Omega_\Psi$
\bea
\label{om_eqs}
&& \frac{\ddot{\Om}_\Psi}{2 \, \Om_\Psi}- \frac{3}{4}
\(\frac{\dot{\Om}_\Psi}{\Om_\Psi}\)^2 +\Om_\Psi^2 = \om_\Psi^2(t), \nonumber \\
&&\dot \varphi_\Psi =\Om_\Psi \, ,\\
&& \Om_\Psi(0)=\om_\Psi(0), ~~~\dot{\Om}_\Psi(0) =0, ~~~ \varphi_\Psi(0)=0, \nonumber
\eea
where dot denotes derivative with respect to time, and where for the future convenience we define the phase $ \varphi_\Psi
\equiv \int_0^t \Om_\Psi(t')dt'  $. 
Taking now into account the initial conditions for the field $\hat\Psi$ itself, we obtain
\be
 \hat\Psi_p(t) = \hat\Psi_p(0^+) \sqrt{\Om_\Psi(0)\over \Om_\Psi(t)} \cos\(\int_0^t \Om_\Psi(t')dt'\)
 +\dot{\hat\Psi}_p(0^+) {1\over \sqrt{\Om_\Psi(0)\Om_\Psi(t)}} \sin\(\int_0^t \Om_\Psi(t')dt'\)\, .
 \labell{fullsol}
\ee
The field $\hat\Psi$ is continuous across the quench,\ie $\hat\Psi_p(0^+)=\hat\Psi_p(0^-)$. The same is true about $\phi$-component of $\dot{\hat\Psi}_p$. However, as argued in section \ref{major}, the time derivative of $\psi$-component exhibits an abrupt jump at $t=0$
\be
 \dot{\hat\psi}_p(0^+)= \dot{\hat\psi}_p(0^-)+i\big(m_\psi(0^+)-\mu_0\big)\hat{\overline\psi}_p(0^-)~.
 \labell{jump}
\ee
Using \reef{fullsol}, it was shown in \cite{Sotiriadis:2010si} that
\be
\label{rho_t}
\rho(t)=\int  \frac{d^2 p}{(2 \pi)^2} \, \frac{1}{2\,\Om_\phi (t)} \left( 1+\frac{\(\Om_\phi(0)-\om_{0p}\)^2}{2 \Om_\phi(0)\om_{0p}} +
\frac{\Om_\phi(0)^2-\om_{0p}^2}{2 \Om_\phi(0)\om_{0p}} \, \cos(2 \varphi_\phi) \right).
\ee
Similarly, using \reef{fullsol}, \reef{jump} and \reef{anticomrel}, the Majorana loop takes the following form
\be
\label{rhopsi_t}
 \rho_\psi(t)=- \int  \frac{d^2 p}{(2 \pi)^2} \, 
 { m_\psi(0^+)\big(p^2+m_\psi(0^+)\mu_0\big)- p^2 \big(m_\psi(0^+)-\mu_0\big)\cos(2\varphi_\psi)\over \Om_\psi(t)\Om_\psi(0)\om_{0p}}~.
\ee

In the limit of large momentum $\Om_\Psi(t)$ approaches $\om_\Psi(t)$ and  therefore both $\rho(t)$ and $\rho_\psi(t)$ exhibit linear divergence. Furthermore, as $p\to\infty$ the oscillatory term in the integrand of  $\rho_\psi(t)$ behaves as $\cos(2 p t)/p$ which upon integration over $p$ boils down to a delta function supported at $t=0$. As argued in section \ref{major} the singularities can be removed using, \eg dimensional regularization scheme. Unfortunately, such a scheme is difficult to implement numerically. Therefore we resort to a different regularization procedure. First, following our discussion in section \ref{major} we separate singularity associated with the delta function
\be
\rho_\psi(t)=\rho_\psi(t)- \big(m_\psi(0^+)-\mu_0\big)\int  \frac{d^2 p}{(2 \pi)^2} {\cos(2pt) \over p}+{m_\psi(0^+)-\mu_0\over 2} \delta(t)~.
\ee
Next we drop $\delta(t)$ since for $t>0$ it vanishes, whereas for
$t=0$ it generates a 
 divergence that can be renormalized away, \ie
\be
\label{rhopsiprime_t}
 \rho_\psi(t)\to\rho'_\psi(t)=\rho_\psi(t) - \big(m_\psi(0^+)-\mu_0\big)\int  \frac{d^2 p}{(2 \pi)^2} {\cos(2pt) \over p}~.
\ee

Finally, we introduce a sharp cut-off $\Lambda$ in the integrals over
momentum to regularize the divergent loops of $\rho'_\psi(t)$ and
$\rho_\phi(t)$.  The terms that depend on the cut-off scale $\Lambda$
can be absorbed in the redefinition of the mass
parameters\footnote{Note that because of delta function singularity,
  mass counterterms at $t=0$ are different from mass counterterms at
  $t>0$. However, such peculiar behaviour of counterterms can be
  attributed to the specific regularization scheme. As we already noticed everything is completely smooth if dimensional regularization is adopted.}. To maintain conformal invariance across the quench instant we set the renormalized mass parameters to zero. As a result, we obtain
\bea
m_{\phi}(t)^2& =& 3 g^2 \( \rho-\frac{\Lam}{4 \pi}\)^2+ g \( \rho'_\psi+ m_\psi(0^+)\frac{\Lam}{2 \pi}\),
\label{mphi}
\\
m_\psi(t)&=&g \( \rho-\frac{\Lam}{4 \pi}\).
\label{meff_reg}
\eea
Together with (\ref{om_eqs},\ref{rho_t}, \ref{rhopsi_t}) and (\ref{rhopsiprime_t})
these give the coupled system of equations that determine the time-evolution of the effective masses. 

We found that for the reasons of numerical stability and accuracy it
is advantageous to rewrite these equations as time-dependent
differential equations for the masses,
\bea
\label{meff_dot}
{d{m}_{\phi}^2\over dt}& =& 6 g^2 \( \rho_\phi-\frac{\Lam}{4 \pi}\) \, \dot{
  \rho}_\phi+g\, \dot{\rho}'_\psi, \nonumber
\\
\dot{m}_\psi&=&g \,\dot{ \rho}_\phi,
\eea
where 
\bea
\label{rhodot}
\dot{\rho}_\phi&=&\int^\Lam_0  dp\, \frac{p}{2 \pi} \(-\frac{\dot{\Om_\phi}}{2
  \,\Om_\phi^2}\) \( 1+\frac{\(\om_\phi(0)-\om_0\)^2}{2 \om_\phi(0)\om_0} +
\frac{\om_\phi(0)^2-\om_0^2}{2 \om_\phi(0)\om_0} \, \cos(2 \varphi_\phi) \) - \nonumber \\
&-& \int^\Lam_0  dp\, \frac{p}{2 \pi} 
\frac{\om_\phi(0)^2-\om_0^2}{2 \om_\phi(0)\om_0} \, \sin(2 \varphi_\phi), \\
\dot\rho'_\psi&=& \int  \frac{d^2 p}{(2 \pi)^2} \,\(\frac{\dot{\Om_\psi}}{\Om_\psi^2}\) \,
 { m_\psi(0^+)\big(p^2+m_\psi(0^+)\mu_0\big)- p^2 \big(m_\psi(0^+)-\mu_0\big)\cos(2\varphi_\psi)\over \Om_\psi(0)\om_{0p}} \non
 &-& 2\big(m_\psi(0^+)-\mu_0\big)\(\int  \frac{d^2 p}{(2 \pi)^2}
  { p^2 \sin(2\varphi_\psi)\over \Om_\psi(0)\om_{0p}}  -\int  \frac{d^2 p}{(2 \pi)^2} \sin(2pt)\)~.
\eea

We solve the above coupled equations numerically by applying a modified version of the algorithm
proposed in \cite{Sotiriadis:2010si}. To this end the equations are
discretized in the momentum space, so that we consider only the
lower uniformly spaced $N_p$ modes with $p \leq \Lambda$. Starting from $t=0$, the set of equations
(\ref{om_eqs}) for the scalar and fermion modes, together with the time-dependent equations for the masses
(\ref{meff_dot}) are advanced in time. In practice, we use the 4th order
Runge-Kutta time-stepping method for the time evolution, and the Simpson
method for evaluation the momentum space integrals in
(\ref{rhodot}), see e.g. \cite{NumRec}. 

Typical values of the numerical parameters used to
generate the results discussed in this paper are in the ranges\footnote{$\mu_0$ is the only dimensionful parameter that sets a scale in our numerical simulations and we choose $\mu_0=1$.} $\Lambda=10-400,~
N_p=2000-50000$, and the discrete time-step is of order $h_t =0.001$. Self-consistency
of the approach requires that effective masses remain well below $\Lambda/4 \pi$, and we
verified that this is indeed satisfied in our case. Other checks
reveal that the results are independent of $N_p$ provided $\Lambda/N_p
\lesssim  0.005$, and the convergence of the method as
a function of $h_t$ is fourth order, meaning that the discretization
errors scale as $O(h_t^4)$ in the continuum limit $h_t \rightarrow 0$.

\begin{figure}[t!]
\hspace{-0.8cm}\includegraphics[width=1.1\textwidth,height=5cm]{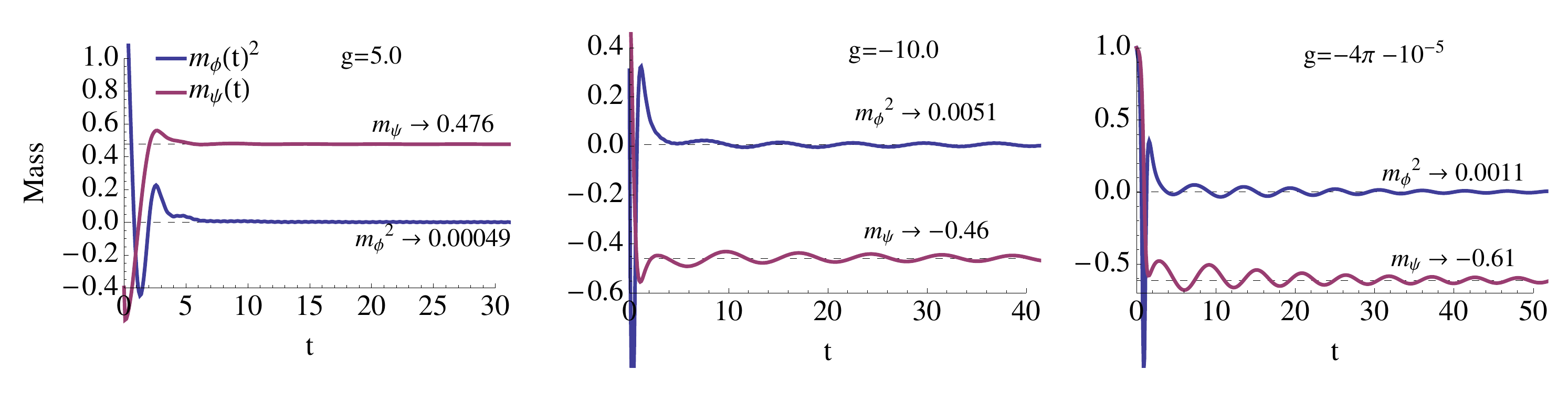}
\caption{Time-evolution
of the effective masses for several values of the coupling $g$. We
find that for generic value of the coupling the effective
masses tend to distinct constants in the late times, signalling
breakdown of supersymmetry. The relaxation time scale grows with $|g|$.
}
\label{fig:meff_gs}
\end{figure}
Figure \ref{fig:meff_gs} shows the temporal dynamics
of the effective masses for several values of the coupling $g$. 
We find that generically the masses approach constant values at late
times, and the relaxation time grows  with $|g|$.  Whilst the
details of the evolution are dictated by the coupling
constant, in all cases the masses of the scalar and the fermion remain distinct,
indicating that supersymmetry is broken by the quench. Typically, the
asymptotic mass of the scalar is very small, being several orders of
magnitude below the mass of the fermion, whose absolute value is of order $0.1$. 

\begin{figure}[t!]
\begin{center}
\includegraphics[width=0.6\textwidth]{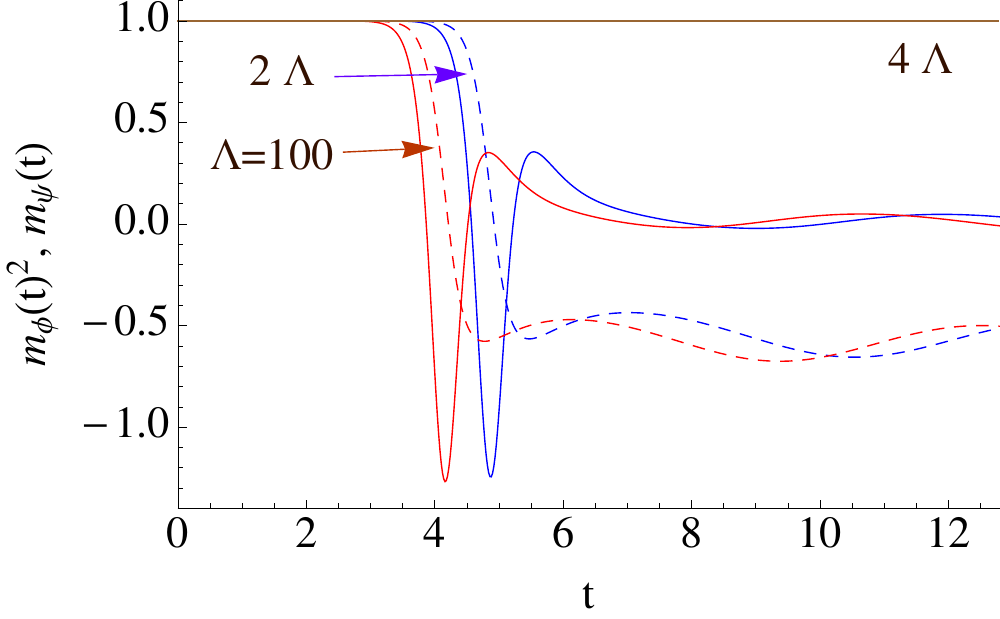}
\caption{The time-evolution
of the effective masses for $g=-4 \pi$ demonstrates instability of
the attractor $m_\psi(t)=m_\phi(t)=\mu_0$ to small numerical errors.}
\label{fig:meff_g4pi}
\end{center}
\end{figure}

As suggested by the analysis of the effective potential in the
previous section in the special case of $g=-4 \pi$ the supersymmetry
is preserved in the quench. In the dynamical setup the effective masses of the scalar and
fermion are initialized as
$m_\psi(0^+)=m_\phi(0^+)=\mu_0=1$. This implies that
$\Omega_\psi=\Omega_\phi \equiv \Omega$. Equations (\ref{rho_t},\ref{rhopsi_t})
yield $\rho_\psi(t)=-2\,\mu_0\,\rho(t) = -2\, \mu_0 \int p\, dp /(2
\pi\, \Omega) $. It follows then from (\ref{mphi},\ref{meff_reg}) that
$m_\psi(t)=m_\phi(t)=\mu_0=1$, are time-independent in the limit of infinite $\Lambda$.
Figure \ref{fig:meff_g4pi} depicts the dynamics of the effective
masses for a set of increasing $\Lambda$'s. It demonstrates that
small numerical imprecisions generated by finite $\Lambda$ eventually
drive the masses away from the initial attractor values. Our numerical method
performs better for larger cut-offs, smaller time-steps etc, such that the masses remain at
their initial values for longer periods of time. The dynamical
evolution at $g=-4 \pi$ and near it (see rightmost panel of
Figure \ref{fig:meff_gs}) indicates that the attractor at
$m_\psi=m_\phi=\mu_0$ is unstable to small perturbations. This is in tune with
the stationary analysis in the last section that showed that in this
case the
effective potential has an inflection point, rather than a minimum,
see Figure \ref{fig:phases}.

\section{Concluding remarks}
\labell{sec:concl}

In this paper we applied the methods proposed in \cite{Sotiriadis:2010si} to quantum quenches in the presence of fermions.  Our findings show that fermionic field responds differently to abrupt changes in the parameters of the theory than its scalar counterpart.  For instance, the qualitative behaviour of fermionic field turns out to be sensitive to the  parity of spacetime dimension  $d+1$, see (\ref{odd_even}). The reason for that can be attributed to the fact that fermions obey the first order Dirac equation, and therefore time derivative of the field experiences a jump at the instant of quench, as opposed to scalars that satisfy the second order Klein-Gordon equation, and therefore both the field and its time derivative are continuous accross the quench.  

To gain a better understanding of this we investigate the expectation value of the fermionic mass term $\langle \bar\psi\psi\rangle$ in the case of free Majorana field. This problem can be solved exactly. We find that for odd dimensional spacetime (even $d$) sharp quenches generate finite and ultra-local terms in $\langle \bar\psi\psi\rangle$,\ie regular functions superposed with delta function or its derivatives supported at the instant of quench, whereas for even dimensional spacetime (odd $d$), sharp quenches result in a singular contribution to $\langle \bar\psi\psi\rangle$ that behaves as $t^{1-d}$ in the vicinity of the quench at $t=0$, but is otherwise  finite everywhere. 

In even $d$'s these singularities can be ignored on account of appropriate regularization scheme accompanied by standard renormalization of the field theory parameters. Indeed, divergences associated with the delta function and its derivatives are scheme dependent and therefore can be removed by a proper choice of regularization scheme, \eg in dimensional regularization they are absent. In odd $d$, however, sharp quenches require a refined analysis since divergences in that case are independent of the choice of regularization scheme. This result is in agreement with \cite{Buchel:2012gw}, where using AdS/CFT it was argued that the limit of sharp quenches is not smooth.

We studied the effect of supersymmetric quantum quenches on the state of the theory at times large compared to any other scale in the problem. No self-consistent computational framework to tackle this problem in general has been found yet. Therefore, we model the quench in supersymmetric $O(N)$ vector model in the limit of large-$N$ and adopt the simplifying assumptions proposed in \cite{Sotiriadis:2010si}. To avoid singularities that may emerge at the instant of sharp quench in even dimensional spacetime and that cannot be removed by a proper choice of regulator, we consider quantum quenches of the simplest supersymmetric extension of the three dimensional ($d=2$) $\phi^6$ model \cite{Hung:2012zr}. 
Using stationary phase approximation we find that supersymmetric quantum quench breaks supersymmetry in the asymptotic state.  

From this perspective the effect of quantum quench is reminiscent of the finite temperature effect. In both cases supersymmetry is broken due to different boundary conditions satisfied by the scalar and fermionic fields. However, SUSY breaking in our case cannot be attributed to thermalization since to leading order in the large $N$ limit the system is integrable, and there is no effective temperature that can be assigned to the final state. 

There is one unfortunate caveat to the above approach. Generic results obtained in this way should be trusted with caution since the Sotiriadis-Cardy approach lacks analytical argument that supports their assumptions. This is why in \cite{Sotiriadis:2010si} they resort to numerical evaluation of exact expressions and eventually find a remarkable match with analytical predictions obtained in the framework of proposed approximation. Note, however, that matching between the analytical and numerical results in \cite{Sotiriadis:2010si} was achieved in the case of scalar field theory only, and it is completely reasonable to expect that behaviour of the fermionic field is very different. 

Therefore, to verify and test our stationary analysis, we derive exact equations of motion and integrate them numerically in time. We find 
that SUSY is broken in the dynamical setting as the system relaxes to a stationary state, see figure \ref{fig:meff_gs}. However, in spite of the qualitative agreement with the analytical predictions, the quantitative details are somewhat different. Perhaps, the chief reason for the apparent discrepancy is intrinsic
assumption in the current approach, that the masses tend to constant values at late times after the initial jump-like transient that leaves no imprint in the asymptotic dynamics. 

This approach proved to be robust in the case of $\phi^4$ field theory in various dimensions. However, as we argued here, fermionic fields exhibit a substantially different response to quantum quenches, and therefore it is not too surprising that some of the assumptions that worked well in the case of scalars are less successful in the case of fermions. In particular, the numerical time-evolution in the case of fermions, shows that the transient is significant even for small $g$'s, and it keeps growing for larger values of the coupling constant. 

In this paper we considered only $O(N)$ symmetric part of the full phase diagram of the model. However, it is of particular interest to explore the impact of quantum quenches on the broken $O(N)$ symmetry, on which some numerical work has appeared in \cite{Gubser}. We leave investigation of this matter for future work.  On the one hand, research in this direction will allow us to compare the full phase diagram of the model in the quenched case with its counterpart at finite temperature \cite{Moshe:2002ra}, while on the other hand, it will provide a better insight into the mechanism of symmetry breaking in general.

\acknowledgments

We would like to thank Alex Buchel and especially Robert C. Myers for useful
discussions and correspondence. Research at Perimeter Institute is
supported by the Government of Canada through Industry Canada and by
the Province of Ontario through the Ministry of Research \&
Innovation. ES is partly
supported by National CITA Fellowship and by NSERC discovery grant.

\appendix

\section{Free Majorana field in 3D}
\label{major2}

In this appendix we present an alternative derivation of eq.\reef{majorprop} that mimics \cite{Sotiriadis:2010si} and exploits the assumption of homogeneous (in space) quench. The latter enables us to study the evolution of each momentum mode $\hat\psi_p(t)$ separately. In the Heisenberg picture, we obtain for $t>0$
 \be
 \ddot{\hat\psi}_p(t)+\omega_p^2\hat\psi_p(t)=0~,
 \ee
 where $\omega_{p}=\sqrt{\vec p^{\,2}+\mu^2}$ , which can be easily solved
  \be
  \hat\psi_p(t)=\hat\psi_p(0^+)\cos(\omega_p t)+\dot{\hat\psi}_p(0^+) {\sin(\omega_p t) \over \omega_p}~.
  \label{majorsol}
  \ee
Recall that $\hat\psi$ is continuous across the instant of quench whereas $\dot{\hat\psi}$ experiences abrupt jump described by eq.\reef{psijump}\footnote{Of course, the same junction condition can be obtained from Dirac equation.}. Therefore
 \bea
 \hat\psi_p(0^+)&=&\hat\psi_p(0^-)~,
 \non
 \dot{\hat\psi}_p(0^+)&=&\dot{\hat\psi}_p(0^-)+i(\mu-\mu_0)\hat{\overline\psi}_p(0^-)~.
 \label{psibc}
 \eea

Using the initial condition that the system lies in the ground state of the hamiltonian which governs the system prior to quench, we get from\reef{MajorFourier}
 \bea
\langle\Psi_0| \hat\psi_{p \al}(0^-) \, \hat\psi_{q \bt}(0^-) |\Psi_0\rangle
&=&(2\pi)^2\delta^{(2)}(p+q){\mu_0 \over \omega_{0p}} u_{0p \al} u^*_{0p \bt} ~, \non
\langle\Psi_0| \hat\psi_{p \al}(0^-) \, \dot{\hat\psi}_{q \bt}(0^-) |\Psi_0\rangle &=&
-\langle\Psi_0| \dot{\hat\psi}_{p \al}(0^-) \, \hat\psi_{q \bt}(0^-) |\Psi_0\rangle
= i (2\pi)^2\delta^{(2)}(p+q) \mu_0 u_{0p \al} u^*_{0p \bt}~, \non
\langle\Psi_0| \dot{\hat\psi}_{p \al}(0^-) \, \dot{\hat\psi}_{q \bt}(0^-) |\Psi_0\rangle
&=&(2\pi)^2\delta^{(2)}(p+q) \mu_0 \, \omega_{0p} \, u_{0p \al} u^*_{0p \bt} ~.
\labell{anticomrel}
 \eea

Combining now eqs.\reef{majorsol} and \reef{psibc} and using the above expectation values yields the same time ordered correlator as in eq.\reef{majorprop}.

\section{Expectation value of the energy momentum tensor}
\labell{emtrace}

In our setup all dimensionfull parameters after the quench are tuned to zero, and the action of the model exhibits conformal invariance. Furthermore, to leading order in $1/N$ all anomalous dimensions vanish. Hence, trace of the energy momentum tensor must vanish as well. We explicitly demonstrate the latter in what follows.

The improved supersymmetric energy momentum tensor of the system after the quench is given by
\bea
T_{\mu\nu}&=&{1\over4} \bar\psi i \big( \gamma_\mu \del_\nu+\gamma_\nu\del_\mu\big)\psi 
+\del_\mu\phi\del_\nu\phi+{1\over 8} (\eta_{\mu\nu}\del^2-\del_\mu\del_\nu)\phi^2
\non
&-&{\eta_{\mu\nu}\over 2}\bigg[   \bar\psi i\ndel\psi -{g\over N}\phi^2(\bar\psi\cdot\psi)
 -2\,{g\over N}(\phi\cdot\bar\psi)(\phi\cdot\psi)
 +\del_\mu\phi\del^\mu\phi-{g^2\over  N^2}(\phi^2)^3 \bigg] \,.
\eea
In particular, to leading order in $1/N$ the expectation value of its trace takes the following form
\be
 \langle T^\mu_\mu\rangle = - \langle  \bar\psi i\ndel\psi \rangle - {1\over 2} \langle \del\phi\del\phi\rangle
 +{3 \, g\over 2N} \langle\phi^2\rangle\langle\bar\psi\psi\rangle + {3\,g^2\over 2 N^2} \langle\phi^2\rangle^3~,
\ee
where we neglected the contribution of $(\phi\cdot\bar\psi)(\phi\cdot\psi)$ since it is of higher order in $1/N$. 

Using now approximation \reef{approx} and eqs.\reef{major2p},\reef{phicorr}, we get as $t\to\infty$
 \be
 \langle  \bar\psi i\ndel\psi \rangle=m_\psi^*\langle\bar\psi\psi\rangle ~, \quad\quad  \langle \del\phi\del\phi\rangle=m_\phi^{*2} \langle\phi^2\rangle~.
 \ee
Hence,
\be
 \langle T^\mu_\mu\rangle =\langle\bar\psi\psi\rangle\( { g\over N} \langle\phi^2\rangle -  m_\psi^* \) 
 -{1\over 2} \langle\phi^2\rangle\( m_\phi^{*2} -{g\over N} \langle\bar\psi\psi\rangle -{3\,g^2\over  N^2} \langle\phi^2\rangle^2\)=0~,
\ee
where the last equality follows from the definition of $m_\psi^*$ and the gap equation \reef{susygap}.

\end{document}